\documentclass{aastex62}

\newcommand\Hzs{Hz s$^{-1}$ }
\newcommand\Hzsns{Hz s$^{-1}$}
\newcommand\totcount{$5\,840\,149$ } 

\newcommand\finalcountnoRFI{$30$ }

\received{November 30, 2018}
\revised{January 11, 2019}

\shorttitle{A search for technosignatures with the GBT}
\shortauthors{Pinchuk et al.}

\begin{document}

\title{A search for technosignatures from TRAPPIST-1, LHS 1140, and 10 planetary systems in the Kepler field with the Green Bank Telescope at 1.15--1.73 GHz
}

\correspondingauthor{Pavlo Pinchuk}
\email{ppinchuk@physics.ucla.edu}

\author[0000-0003-4736-4728]{Pavlo Pinchuk}
\affil{Department of Physics and Astronomy, University of California, Los Angeles, CA 90095, USA}

\author[0000-0001-9798-1797]{Jean-Luc Margot}
\affiliation{Department of Physics and Astronomy, University of California, Los Angeles, CA 90095, USA}
\affiliation{Department of Earth, Planetary, and Space Sciences, University of California, Los Angeles, CA 90095, USA}

\author[0000-0001-8834-9423]{Adam H. Greenberg} 
\affiliation{Department of Physics and Astronomy, University of California, Los Angeles, CA 90095, USA}

\author{Thomas Ayalde} 
\affiliation{Department of Physics and Astronomy, University of California, Los Angeles, CA 90095, USA}

\author{Chad Bloxham} 
\affiliation{Department of Physics and Astronomy, University of California, Los Angeles, CA 90095, USA}

\author{Arjun Boddu} 
\affiliation{Department of Electrical Engineering, University of California, Los Angeles, CA 90095, USA}

\author{Luis Gerardo Chinchilla-Garcia} 
\affiliation{Department of Physics and Astronomy, University of California, Los Angeles, CA 90095, USA}

\author{Micah Cliffe} 
\affiliation{Department of Electrical Engineering, University of California, Los Angeles, CA 90095, USA}

\author{Sara Gallagher} 
\affiliation{Department of Physics and Astronomy, University of California, Los Angeles, CA 90095, USA}

\author{Kira Hart} 
\affiliation{Department of Physics and Astronomy, University of California, Los Angeles, CA 90095, USA}

\author{Brayden Hesford} 
\affiliation{Department of Electrical Engineering, University of California, Los Angeles, CA 90095, USA}

\author{Inbal Mizrahi} 
\affiliation{Department of Physics and Astronomy, University of California, Los Angeles, CA 90095, USA}

\author{Ruth Pike} 
\affiliation{Department of Earth Science and Engineering, Imperial College London, UK}

\author{Dominic Rodger} 
\affiliation{Department of Earth Science and Engineering, Imperial College London, UK}

\author{Bade Sayki} 
\affiliation{Department of Physics and Astronomy, University of California, Los Angeles, CA 90095, USA}

\author{Una Schneck} 
\affiliation{Department of Earth, Planetary, and Space Sciences, University of California, Los Angeles, CA 90095, USA}

\author{Aysen Tan} 
\affiliation{Department of Electrical Engineering, University of California, Los Angeles, CA 90095, USA}

\author{Yinxue ``Yolanda'' Xiao} 
\affiliation{Department of Computer Science, University of California, Los Angeles, CA 90095, USA}

\author{Ryan S.\ Lynch}
\affiliation{Green Bank Observatory, PO Box 2, Green Bank, WV
  24494, USA}
\affiliation{Center for Gravitational Waves and Cosmology, Department
  of Physics and Astronomy, West Virginia University, White Hall, Box
  6315, Morgantown, WV 26506, USA}

\begin{abstract}
  As part of our ongoing search for technosignatures, we collected over three terabytes of data in May 2017 with the L-band receiver (1.15--1.73 GHz) of the 100 m diameter Green Bank Telescope. These observations focused primarily on planetary systems in the Kepler field, but also included scans of the recently discovered TRAPPIST-1 and LHS 1140 systems. We present the results of our search for narrowband signals in this data set with techniques that are generally similar to those described by \citet{Margot2018}. Our improved data processing pipeline classified over $98\%$ of the $\sim$ 6 million detected signals as anthropogenic Radio Frequency Interference (RFI). Of the remaining candidates, \finalcountnoRFI were detected outside of densely populated frequency regions attributable to RFI. These candidates were carefully examined and determined to be of terrestrial origin. We discuss the problems associated with the common practice of ignoring frequency space around candidate detections in radio technosignature detection pipelines. These problems include inaccurate estimates of figures of merit and unreliable upper limits on the prevalence of technosignatures. We present an algorithm that mitigates these problems and improves the efficiency of the search. Specifically, our new algorithm increases the number of candidate detections by a factor of more than four compared to \citet{Margot2018}.
\end{abstract}


\section{Introduction} \label{sec:intro}

The question ``Are we alone in the Universe?'' is one of the most enduring and fundamental unanswered questions in modern science. The quest for a definitive answer to this question motivates the search for evidence of life in the Solar System and beyond. Over time, two primary strategies for the search have emerged. One focuses on biosignatures, which are defined as scientific evidence of past or present life. The other focuses on technosignatures, which are defined as any scientific evidence of the existence of 
\added{extinct or extant} technology. Given our present knowledge of astrobiology, it is impossible to reliably predict which strategy will succeed first. We argue that the search for technosignatures offers four advantages compared to telescopic or robotic searches for biosignatures.

First, present-day human technology limits the targets for a biosignature search to planetary systems around a few dozen nearby stars, including our own Sun. In contrast, the number of available targets for a directed technosignature search is likely at least a billion times larger, given the abundance of potentially habitable exoplanets \citep{Borucki} and our ability to detect technosignatures emitted thousands of light years away \citep[e.g.,][]{Margot2018}.
Second, the detection of a biosignature
outside the Solar System, along with its interpretation, may remain ambiguous and
controversial for many years, whereas a confirmed detection of an extraterrestrial technosignature 
\added{of the type described in this work}
would offer a high level of certainty in interpretation.
Third, the information content of some biosignatures
may remain
limited post-detection, while the content of an intentional extraterrestrial transmission
could potentially provide an
unparalleled advance in knowledge. Finally, 
we estimate that
a substantial search for technosignatures can be accomplished with only a
small ($<$1\%) fraction of the budget currently allocated to the search for biosignatures,
\added{including missions like Mars 2020, Europa Clipper, and the James Webb Space Telescope}.

In this work, we present a search for radio technosignatures using the L-band receiver of the
100 m diameter Green Bank Telescope (GBT). We scanned a total of 12 sources within a 2 hour observational window
with the goal of detecting narrowband emissions. Narrowband ($\sim10$ Hz) signals are diagnostic of engineered emitters because the narrowest known sources of natural emission span a larger ($\sim500$ Hz) range of frequencies at L band \citep[e.g.,][]{cohe87}.  
Our search builds on the legacy of technosignature searches performed in the
period 1960--2010 \citep[][and references therein]{Tarter2001, Tarter2010}
and  complements some recent efforts \citep{Siemion2013, Harp, Enriquez2017, Gray, Margot2018}, but differs primarily in the choice of sample of the vast parameter space still left to search. In addition, we are generally sensitive to a wider range of signal drift rates ($\pm8.86$ \Hzsns) than the cited works, and to signals with lower signal-to-noise-ratio (SNR$>$10) 
than a recent large survey \citep[][SNR$>$25]{Enriquez2017}.

We present a substantial improvement to the
signal detection algorithms used
in the
search pipelines of \citet{Siemion2013, Enriquez2017, Margot2018}.
In 
these pipelines, 
a wide (several hundred Hertz) window surrounding the frequency of each candidate detection is removed from further consideration, whether this candidate appears in a primary \citep{Siemion2013, Margot2018} or secondary \citep{Enriquez2017} scan.

Consequently, other legitimate signals within that window are never analyzed, which implies that both figures of merit for the completeness of the search and upper limits on the prevalence of technosignatures reported in these works are inaccurate. By adjusting the way in which the frequency space surrounding each candidate signal is handled, we no longer need to discard legitimate signals and are able to analyze a larger fraction of the frequency space.
This advance is appreciable. We estimate that this improvement alone increases our detection count by a factor of more than four as compared to \citet{Margot2018}. The increase in the detection count of other search pipelines after implementation of a similar improvement would likely be substantial.

Our data acquisition techniques, a brief overview of which is presented in Section \ref{sec:data aq}, are generally similar to those presented by \citet{Margot2018}.
Section \ref{sec:analysis} explains our data analysis procedures, including major changes and improvements to our data processing pipeline. The results of our search are presented in Section \ref{sec:results}, followed by a discussion and conclusions in Sections \ref{sec:discussion} and \ref{sec:conclusions}, respectively. 

\section{Data Acquisition} \label{sec:data aq}

\subsection{Sources} \label{subsec:sources}

For our observations, we selected 10 sources (Table \ref{tab:soucres}) from the Kepler catalog based on habitability criteria presented by \citet{Kane}. These criteria take into account the size and location of exoplanets with respect to the host stars. Small 
planets ($R_p < 2R_E$, where $R_p$ and $R_E$ are the radii of the planet and Earth, respectively)
in the conservative and optimistic habitable zones fall under Habitable Zone (HZ) categories 1 and 2, respectively. Conversely, planets with any radius in the conservative and optimistic habitable zones fall under HZ categories 3 and 4, respectively. 

For our observations, we selected all seven host stars reported by \citet{Kane} with confirmed planets in HZ categories 1 and 2 (Table \ref{tab:soucres}). We supplemented these with three more host stars with confirmed planets in HZ categories 3 and 4 (Table \ref{tab:soucres}).
In addition to the 10 sources in the Kepler field, we observed the recently discovered TRAPPIST-1 and \replaced{LHS 110}{LHS 1140} systems. The TRAPPIST-1 system hosts seven Earth-sized, temperate exoplanets orbiting an ultra-cool dwarf star \citep{TRAPPIST1}. The benign equilibrium temperatures of some of the planets in the system make the prospect of liquid water on their surfaces, and thus the possibility of life, plausible. The LHS 1140 system
harbors only one known planet, which orbits its M dwarf host star within the
habitable zone \citep{LHS1140}.

\begin{deluxetable}{lcCcc}
  \caption{Target host stars listed in order of observation. Distances
    in light years (ly) were obtained from the NASA Exoplanet Archive. Habitable Zone categories are described by \citet{Kane}. Categories 1 and 2 refer to small $(R_p < 2R_E)$ planets in the conservative and optimistic habitable zones, respectively, while categories 3 and 4 refer to planets of any radius in the conservative and optimistic habitable zones, respectively. TRAPPIST-1 and LHS 1140 were categorized
    on the basis of orbital radii from the NASA Exoplanet Archive and HZ boundaries as calculated with the algorithm of \citep{Kopp}.
    \label{tab:soucres}}
\tablehead{
\colhead{Host Star} & \colhead{} & \colhead{Distance (ly)} & \colhead{} & \colhead{HZ 
Category}}
\startdata
Kepler-442 && $1115^{+62}_{-72}$		&&  1, 2 \\ 
Kepler-440 && $851^{+52}_{-150}$      	&&  2    \\ 
Kepler-174 && $1210.04^{+63}_{-56}$ 	&&  3, 4 \\ 
Kepler-62  && $1200$      				&&  1, 2 \\ 
Kepler-296 && $737^{+91}_{-59}$      	&&  1, 2 \\ 
Kepler-86  && $1128.5^{+44}_{-42}$  	&&  4    \\ 
Kepler-22  && $620$      				&&  4    \\ 
Kepler-283 && $1477.49^{+67}_{-74}$ 	&&  1, 2 \\ 
Kepler-452 && $1787.34^{+395}_{-243}$  	&&  2    \\ 
Kepler-186 && $561^{+42}_{-33}$      	&&  1, 2 \\ 
TRAPPIST-1 && $39.5\pm1.3$      		&&  1, 2 \\ 
LHS 1140   && $40.67\pm1.37$      		&&  1    \\ 
\enddata
\end{deluxetable}

\subsection{Observations} \label{subsec:obs}

We conducted our observations with the 100 m diameter Green Bank Telescope (GBT) on May 4, 2017, 15:00 – 17:00 Universal time (UT). We recorded both linear polarizations of the L-band receiver using the GUPPI backend in its baseband recording mode \citep{GUPPI}. GUPPI was configured to channelize 800 MHz of recorded bandwidth into 256 channels of 3.125 MHz each.  
We verified telescope pointing accuracy with standard procedures and tested signal integrity by injecting a monochromatic tone near the receiver frontend.

We observed all our targets in pairs in order to facilitate the detection and removal of signals of terrestrial origin (Section \ref{subsec:SQL_filters}). The sources were paired by
approximately minimizing the slew time of the telescope for the duration of the observing block and then taking consecutive pairs from the solution. Pairings were adjusted to eliminate any ambiguity in the direction of origin of detected signals. Specifically, we required angular separations larger than $1^{\circ}$ (several times the 
$\sim9$ arcmin
beamwidth of the GBT
at the center frequency of our observations) 
between pair members. The pairs are listed consecutively in Table \ref{tab:soucres} (i.e., the first pair is Kepler-442 and Kepler-440, and so on). Each pair was observed for a total of $\sim330$ s using an ``on'' - ``off'' - ``on'' - ``off'' sequence, where ``on'' represents a scan of the first source in the pair, and ``off'' represents a scan of its partner.

\section{Analysis} \label{sec:analysis}

\subsection{Data Pre-Processing} \label{subsec:preproc}
After unpacking the data, we computed $2^{20}$-point Fourier transforms of the digitized 
raw voltages, yielding a frequency resolution of 
$\Delta \nu = 2.98$ Hz.
We chose this frequency resolution because it is small enough to provide unambiguous detections of technosignatures and large enough to examine Doppler frequency drift rates of up to $\sim10$ Hz s$^{-1}$ (Section \ref{subsec:treealg}).

We modeled the bandpass response of GUPPI's 256 channels by fitting a 16-degree Chebyshev polynomial to the median bandpass response of all channels within the operating range of the GBT L-band receiver (1.15 –- 1.73 GHz), excluding channels that overlap the frequency range 
(1200 -- 1341.2 MHz) 
of a notch filter designed to mitigate radio frequency interference (RFI) from aircraft radar.  Our experience indicates that this process allows us to apply bandpass corrections that yield the expected flat baselines. After applying the bandpass correction to all 256 channels, we stored consecutive power spectra of length $2^{20}$ as rows in a
time-frequency array and normalized the result to zero mean and unit variance of the noise power. We call the graphical representation of such arrays time-frequency diagrams, \added{though they are also often referred to as ``spectrograms'', ``spectral waterfalls'', or ``waterfall plots''}.

\subsection{Doppler De-Smearing} \label{subsec:treealg}
Due to the orbital and rotational motions of both the emitter and the receiver, we expect extraterrestrial technosignatures to drift in frequency space \citep[e.g.,][]{Siemion2013, Margot2018}. A de-smearing algorithm is required to avoid spreading the power of a given signal over multiple channels. Since the Doppler drift rates due to the emitters are unknown, we examined 1023 linearly spaced drift rates in the range $\pm8.86$ \Hzsns, with a step of 
$\Delta \dot{f} = 0.0173$ \Hzsns. 
To accomplish this, we made use of a computationally advantageous Doppler de-smearing algorithm \citep{Taylor1974, Siemion2013} which computes an array containing de-smeared power spectra, where each de-smeared spectrum represents a time integration of the consecutive power spectra after correcting for a given Doppler drift rate. A single pass of this algorithm computes 512 power spectra for all drift rates in the range 0 to 8.86 \Hzs. To obtain power spectra at negative drift rates, we applied the algorithm a second time and reversed the search direction. 

\begin{figure}
\plotone{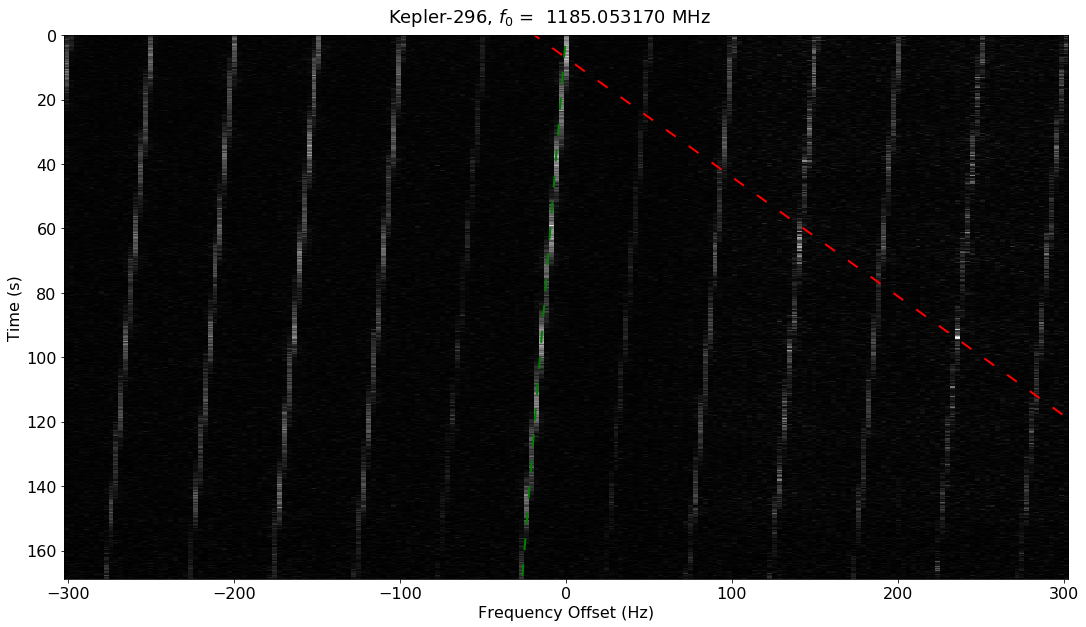}
\caption{Time-Frequency diagram of a signal that would result in multiple detections in an earlier version of our data processing pipeline. In this diagram, pixel intensity represents signal power. The green dashed line shows the correct drift rate of the signal, while the red dashed line indicates a drift rate along which signal values integrate to a summed power that exceeds our detection threshold, resulting in an additional, spurious detection. Our improved pipeline eliminates the possibility of spurious detections for signals of this nature.\label{fig:f1}}
\end{figure}

In a previous analysis, \citet{Margot2018} performed a search for technosignatures on each of the resulting arrays individually.  As a result, it was possible for a candidate signal to be detected twice; once with the correct drift rate, and once with a spurious drift rate of the opposite sign (Figure \ref{fig:f1}).
This duplication increased the false detection count and was occasionally problematic in subsequent stages of the data-processing pipeline. To avoid this problem, we concatenated the outputs of both applications of the de-smearing algorithm into a single array  prior to subsequent analysis. This array contained all 1023 possible drift rates (one duplicate calculation at 0 \Hzs was removed).

\subsection{Candidate Signal Detection} \label{subsec:blanking}

The output of the Doppler de-smearing algorithm contains the integrated power spectra of the scan at various drift rates, making it ideal for identifying promising candidates for extraterrestrial technosignatures (hereafter, ``candidate signals''). We performed this search iteratively by identifying the signal with the highest SNR and storing its characteristics in a
structured query language (SQL) database before moving on to the next signal. Drift rates that are similar to the drift rate with maximum SNR often yield integrated powers with large SNR values as well, and it is important to avoid redundant detections of the same signal.  
\citet{Siemion2013, Margot2018} avoided the redundancy by discarding all detections within a frequency range $\Delta f$ centered on the start frequency of the highest-SNR signal.
For the analysis at 3 Hz resolution, they defined the frequency range to be $\Delta f = 2 \dot{f}_{max}\tau$, where $\dot{f}_{max} =$ 8.86 \Hzs and $\tau$ is the 
duration of the scan (typically around $150$ s). This choice guaranteed that
duplicate detections for signals with the highest detectable drift rates were not recorded.
However, this procedure also removed all other valid candidates within a
$\sim3000$ Hz window of every detected candidate signal. Moreover, due to the iterative nature of the search, high SNR signals were always detected first,
which prevented lower-SNR signals in their vicinity from being detected.   
This procedure removed many legitimate candidate signals in the vicinity of
higher-SNR signals, ultimately leaving large regions of the spectrum unexamined.
This issue results in two unacceptable consequences.  First, it yields incorrect calculations of figures of merit, because such calculations rely on an accurate measure of the bandwidth examined during a particular search.  Second, it renders attempts to place upper limits on the abundance of technosignature sources unreliable because the pipeline eliminates the very signals it purports to detect.  Problems of this nature affect the results of several studies, including those of \citet{Siemion2013, Enriquez2017, Margot2018}.
To properly quantify the impact of these discarded signals on estimates of the prevalence of technosignatures, a proper injection and recovery study must be conducted.  This study is beyond the scope of our work.

We designed a novel procedure to alleviate these shortcomings.
In order to avoid redundant detections, we simply require that no new detection contain any subset of the points in time-frequency space belonging to any other already-detected signal. In other words, we require that none of our detections ``cross'' in time-frequency space. 
In this context, the frequency range used to discard redundant detections is no longer a constant, but rather a function of the drift rate of a new potential candidate and the bandwidth of the already-detected signal.
If $\dot f_0$ is the drift rate of a known candidate, and $\dot f$ is the drift rate of the potential candidate signal, then the potential candidate is marked redundant if its frequency $f$ at the start of the scan satisfies
\begin{eqnarray} \label{eq:blanking}
f_0 - \Delta f_b  <  f <  f_0 + (\dot f - \dot f_0 ) \tau + \Delta f_b \qquad \text{if} \quad \dot f > \dot f_0 \\
\nonumber
f_0 + \Delta f_b  >  f >  f_0 + (\dot f - \dot f_0 ) \tau - \Delta f_b  \qquad \text{if} \quad \dot f < \dot f_0
\end{eqnarray}
where $f_0$ is the frequency of a known candidate signal, $\tau$ is the scan
duration, and $\Delta f_b$ is half of the
signal bandwidth.
Because we do not want new detections
that contain \textit{any} part of an already-detected signal, we must account for
its non-zero bandwidth by extending the frequency range as in Equation \ref{eq:blanking}. For implementation details of this procedure, including the estimation of bandwidth, see Appendix \ref{sec:bw}.

One
drawback of this method is that potential technosignatures may be discarded if they ``cross''
a stronger signal of terrestrial origin in time-frequency space. There is no reason to assume \textit{a priori} that a valid candidate signal would not exhibit this behavior.
However, superimposed signals are, by their very nature, difficult to detect.
Other detection pipelines, including those of \citet{Siemion2013, Enriquez2017, Margot2018} are also blind to such signals.

We identified all candidate signals with SNR
$> 10$ in channels within the operating range of the GBT L-band receiver (1.15 –- 1.73 GHz), excluding channels that overlap the GBT notch filter at 1200 -- 1341.2 MHz.  A total of  \totcount candidate signals were detected.

\subsection{Doppler and Direction-of-Origin Filters} \label{subsec:SQL_filters}
After all candidate signals were identified, we applied several filter procedures
in order to distinguish anthropogenic signals from promising technosignature candidates. The overall purpose of the algorithms is generally similar to those described by \citet{Margot2018}.  
\deleted{For implementation details, see Appendix \ref{sec:SQLfilters}.}

Our filter procedures (hereafter, ``filters'') are designed to search the SQL database and flag the most promising candidates. The first filter flagged all candidate signals with non-zero Doppler drift rates. Signals  
with zero Doppler drift rates\added{, defined here as signals that drift across less than one frequency channel over the course of a scan,}
are of no interest to us
because the corresponding emitters will generally not be in motion with respect to the receiver, which suggests they are terrestrial in nature. The second filter flagged a signal as a promising candidate if it is persistent, i.e., it
is detected in both scans of its source.
This filter removes any anthropogenic signals that may have temporarily entered the beam during
one of the scans. The third filter marks a signal as a technosignature candidate if
its direction of origin is unique, i.e, it
is not detected in the scans corresponding to other sources.
This filter eliminates many anthropogenic signals that are detectable through the antenna sidelobes. 
\added{For implementation details, see Appendix \ref{sec:SQLfilters}.}
  
If both a candidate signal and its
corresponding signal in the other scan of the source were flagged by all three filters, then the candidate signal was marked as a high-interest signal. Of the \totcount total detections, 
$5\,743\,209$ ($>$ 98\%)
were discarded as a result of these filters.

\subsection{Frequency Filters} \label{subsec:Freq_filters}
\begin{figure}[ht!]
\includegraphics[width=\textwidth]{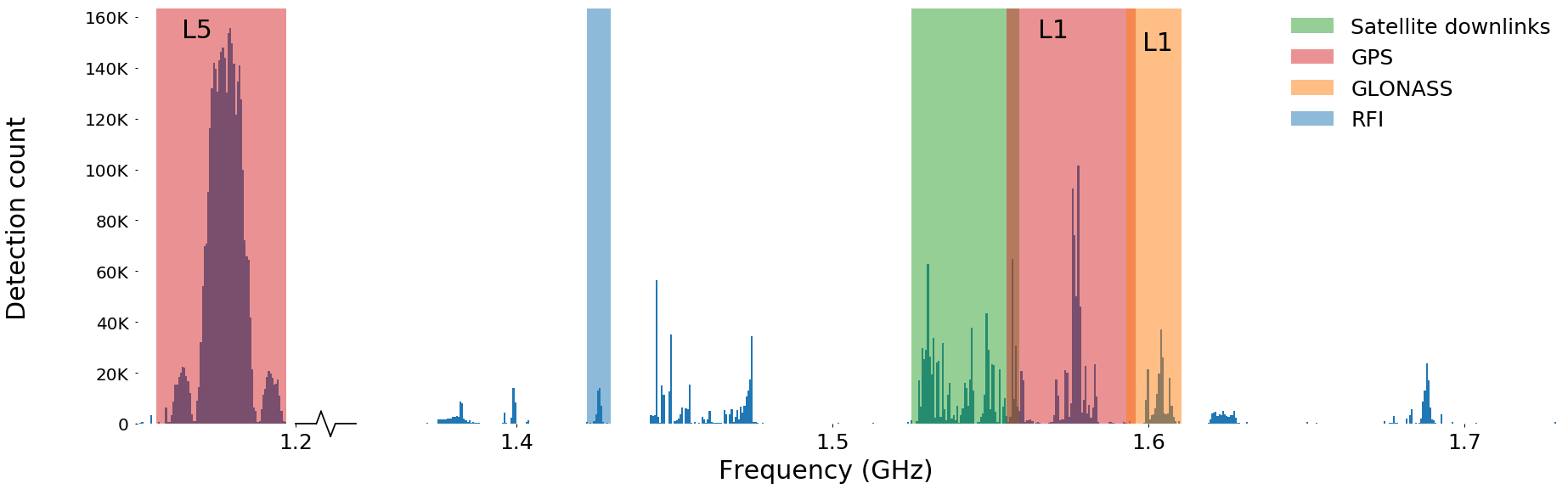}
\caption{Candidate signal count as a function of frequency, superimposed on operating bands of known interferers. 
Note that the majority of detections occur in congested bands. The region marked simply as ``RFI'' contains signals from an interferer, likely ARSR, 
  whose intermodulation products overlap the radio astronomy protected band \citep{Margot2018}.  Frequencies overlapping the GBT notch filter at 1200 -- 1341.2 MHz are excluded. 
  \label{fig:cand_count}
}       
\end{figure}

A majority of the signals detected in our search have frequencies in operating bands of known interferers, such as Global Navigation Satellite Systems or Aircraft Radar. Figure \ref{fig:cand_count} depicts our detection count superimposed onto frequency bands of known anthropogenic RFI. The interferer labeled simply as ``RFI'' is particularly interesting, as it overlaps the radio astronomy protected band near 1420 MHz. The time-frequency structure of this RFI is similar to that described by \citet{Siemion2013} and \citet{Margot2018}, who attributed the likely origin of the  RFI to intermodulation products of Air Route Surveillance Radars (ARSR). Table  \ref{tab:known_RFI} describes the properties of the regions of most prominent detected anthropogenic RFI.
These regions are reminiscent of some of \citet[][Table 2]{Harp}'s ``permanent RFI bands.''
All candidate signals detected within these regions were removed from consideration because of their likely anthropogenic nature.

\begin{deluxetable}{lcCc}[ht!]
\caption{Spectral regions exhibiting a high density of detections per unit frequency. Known anthropogenic interferers are listed in the `Identification' column.
\label{tab:known_RFI}}
\tablehead{
\colhead{Frequency Region (MHz)} & \colhead{Total detection count} & \colhead{Density (\# per MHz)} & \colhead{Identification}}
\startdata
1155.99 - 1196.91    & $3\,579\,122$ & $87\,466$ &  GPS L5              \\
1192.02 - 1212.48    & $96\,233$     & 4703      &  GLONASS L3          \\
1422.320 - 1429.992  & $41\,757$     & 5443      &  ARSR products?      \\ 
1525 - 1559          & $809\,877$    & $23\,820$ &  Satellite downlinks \\
1554.96 - 1595.88    & $718\,711$    & $17\,564$ &  GPS L1              \\
1592.9525 - 1610.485 & $184\,207$    & $10\,507$ &  GLONASS L1          \\
\enddata
\end{deluxetable}

\begin{figure}[ht!]
  \plotone{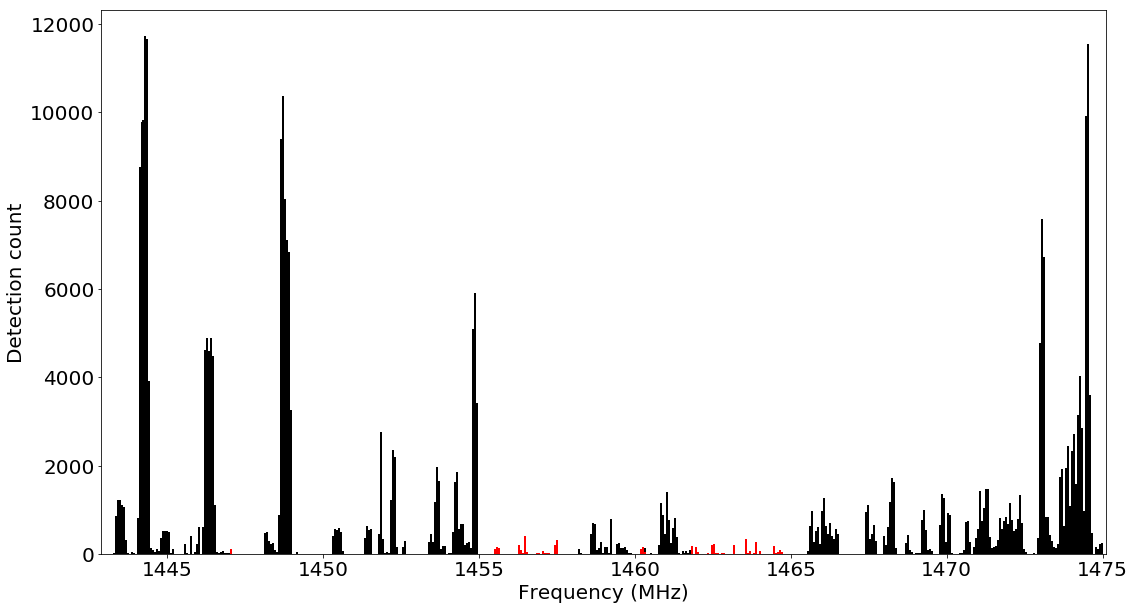}
\caption{Example of a region 
of high signal density detected outside of known interferer operating bands. Signals left after
  application of our
  signal density thresholding procedure are shown in red.
  Most of the remaining signals are eliminated by the Doppler and direction-of-origin filters, which are applied independently.  
  \label{fig:DenseRFI}}
\end{figure}

After removing candidates 
with frequencies in the operating bands of some well-known anthropogenic interferers (Table~\ref{tab:known_RFI}),
we observed that
regions of high 
signal density remained. An example of such a region is shown in Figure \ref{fig:DenseRFI}. Note the presence of many narrow clusters of high-signal-density regions.
Although it is possible that a valid technosignature could be found within one of these regions, it would likely be difficult to detect and  establish its validity given the vast amount of strong, ambient RFI.
With this challenge in mind, we developed a method to discriminate signals found within densely populated frequency regions from signals found in quieter parts of the spectrum.
Specifically, we measured the signal density in 1-kHz-wide frequency bins and
discarded candidates within that bin if the signal density exceeded a pre-defined threshold value. For this work, we chose a threshold value of 1000 signals MHz$^{-1}$,
which corresponds to a sharp drop-off in histograms of signal density.  
For further details, see Appendix \ref{sec:DensityThresh}. A sample result of this procedure is shown in Figure \ref{fig:DenseRFI}, where signals left after our density thresholding procedure are shown in red. 
More than 96\% of the 581\,433 signals found outside of the well-known interferer operating bands listed in Table \ref{tab:known_RFI} were discarded using this procedure.

\section{Results}\label{sec:results}

\begin{deluxetable}{lcCccc}
\caption{Characteristics of Top 30 Candidates
\label{tab:final_cands}}
\tablehead{
\colhead{Source} & \colhead{Epoch (MJD)}  & \colhead{Frequency (Hz)} & \colhead{Drift Rate \Hzs} & \colhead{SNR} & \colhead{Category}}
\startdata
 Kepler-86  & 57877.66340278 & 1151559391.617775 &  0.0867  &  11.2 & a \\ 
            &                & 1457412120.699883 &  0.4510  &  19.7 & b \\ 
\hline 
 Kepler-174 & 57877.65087963 & 1501593533.158302 & -0.0520  &  33.8 & b \\ 
\hline 
 Kepler-186 & 57877.68348380 & 1457412961.125374 &  0.5551  &  18.6 & c \\ 
            &                & 1457488715.648651 &  0.1908  &  15.6 & a \\ 
            &                & 1457489272.952080 &  0.1735  &  12.9 & a \\ 
            &                & 1693601790.070534 &  0.1041  & 189.0 & b \\ 
\hline
 Kepler-296 & 57877.66091435 & 1420354264.974594 &  0.2429  &  11.2 & f \\
            &                & 1420476266.741753 &  0.2255  &  15.4 & f \\
            &                & 1431607288.122177 &  0.2082  &  10.5 & f \\
            &                & 1435340115.427971 &  0.1561  &  26.6 & a \\ 
\hline 
 Kepler-440 & 57877.64343750 & 1457473185.658455 &  0.3296  &  26.2 & b \\ 
\hline 
 Kepler-442 & 57877.64098380 & 1436369919.776917 &  0.3123  &  11.3 & a \\ 
            &                & 1457510107.755661 &  0.2949  &  22.4 & d \\ 
            &                & 1501766377.687454 & -0.0520  &  17.8 & e \\ 
\hline
 LHS 1140   & 57877.69550926 & 1675002533.197403 &  0.01735 &  28.0 & f \\
            &                & 1676205691.695213 &  0.01735 &  23.5 & f \\
            &                & 1676211157.441139 & -0.01735 &  24.0 & f \\
            &                & 1677041041.851044 & -0.01735 &  12.9 & f \\
            &                & 1678640750.050545 &  0.03469 &  17.0 & f \\
            &                & 1728618854.284286 & -0.01735 &  10.0 &   \\
\hline 
 TRAPPIST-1 & 57877.69228009 & 1151731526.851654 &  0.01735 &  16.2 &   \\
            &                & 1463927415.013313 &  0.03469 &  92.4 & a \\ 
            &                & 1675910618.901253 & -0.01735 &  14.0 & f \\
            &                & 1676249122.619629 & -0.01735 &  13.1 & f \\
            &                & 1676298260.688782 & -0.01735 &  14.6 & f \\
            &                & 1677018946.409225 & -0.01735 &  14.1 & f \\
            &                & 1678110888.600349 &  0.03469 &  12.0 & f \\
            &                & 1678460314.869881 &  0.01735 &  13.6 & f \\
            &                & 1678716921.806335 &  0.01735 &  15.5 & f \\
\enddata
\tablecomments{Properties are listed for the first scan of the source only. For a description of the `Category' column, see Section \ref{sec:results}.}
\end{deluxetable}

Signals that remained after application of our
Doppler, direction-of-origin, and frequency filters
were marked as final technosignature candidates. Thirty such signals remained, and their properties are given in Table \ref{tab:final_cands}.
Further examination of the final 30 candidates revealed that 13 of them
are anthropogenic because they are also present in the ``off'' scan of the source.
They were not correctly identified by our filters for a variety of reasons, which we summarize into categories below. 

Category `a' refers to signals with 
SNR$<$10 in the ``off'' scan.
Because the ``off'' scan detections were not recorded in the database, it was not possible for our filters to flag this category of signals as RFI. 

Category `b' refers to signals whose drift rates between the ``on'' and ``off'' scans differed by more than our allowed tolerance, which we set to  $\pm\Delta \dot{f}$ = $\pm0.0173$ \Hzs (Appendix \ref{sec:SQLfilters}).
Because the ``off'' scan detections were not correctly paired to the ``on'' scan detections, it was not possible for our filters to flag this category of signals as RFI.

Category `c' refers to signals that `cross' a signal of a higher SNR in scan of a different source. As discussed before, our signal detection methods are currently blind to such signals.

Category `d' refers to broad signals for which it is difficult to accurately determine a drift rate. Such signals are naturally difficult to `pair', since the integrated power may peak at different frequencies
within the bandwidth of the signal 
for different scans. 

Category `e' refers to signals that exhibit non-linear behavior in frequency as a function of time. Our pipeline is not currently well-equipped to handle such signals because the 
Doppler de-smearing
algorithm can only detect linear drifts in frequency vs. time.  

All categories described above pinpoint potential areas for improvement for our current pipeline.

\begin{figure*}
\gridline{\fig{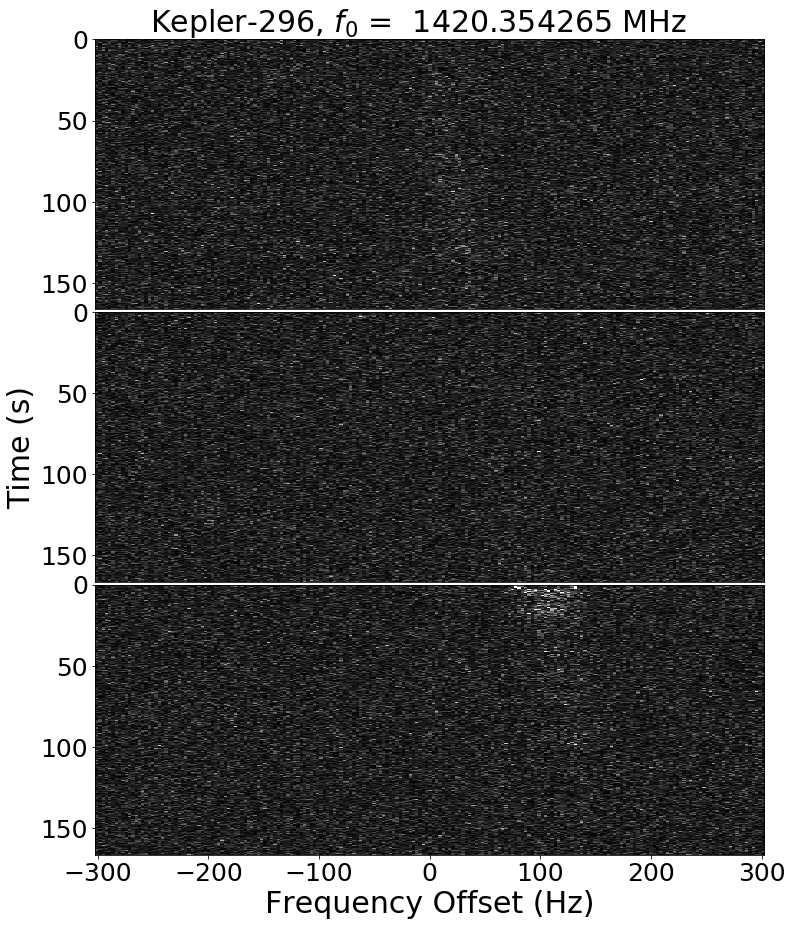}{0.3\textwidth}{}
          \fig{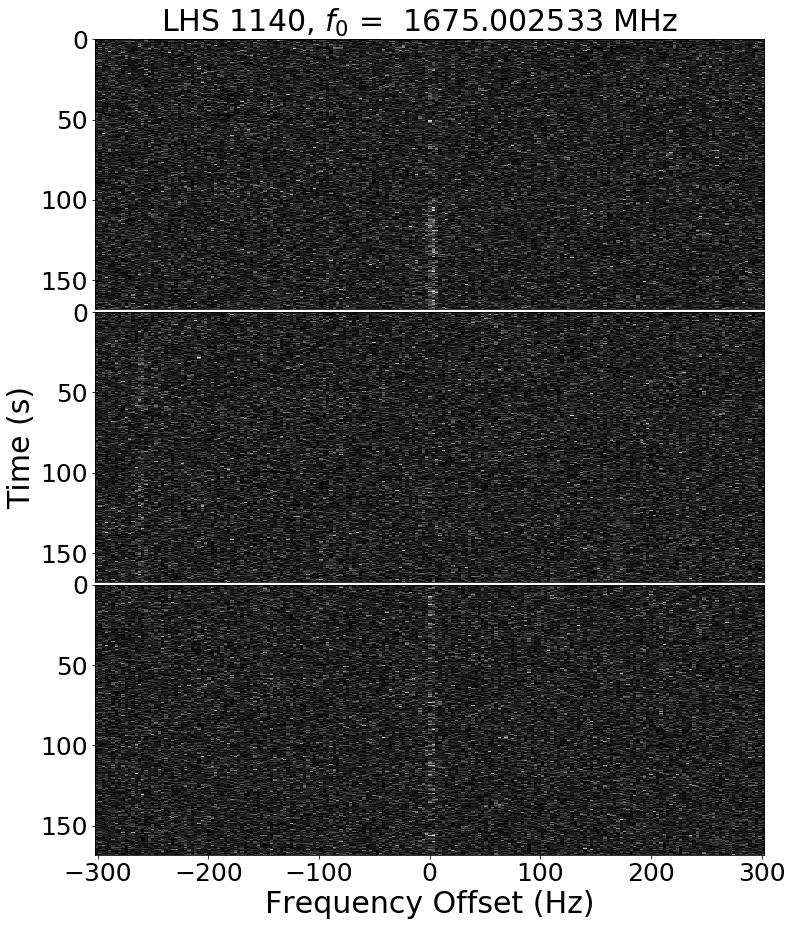}{0.3\textwidth}{}
          \fig{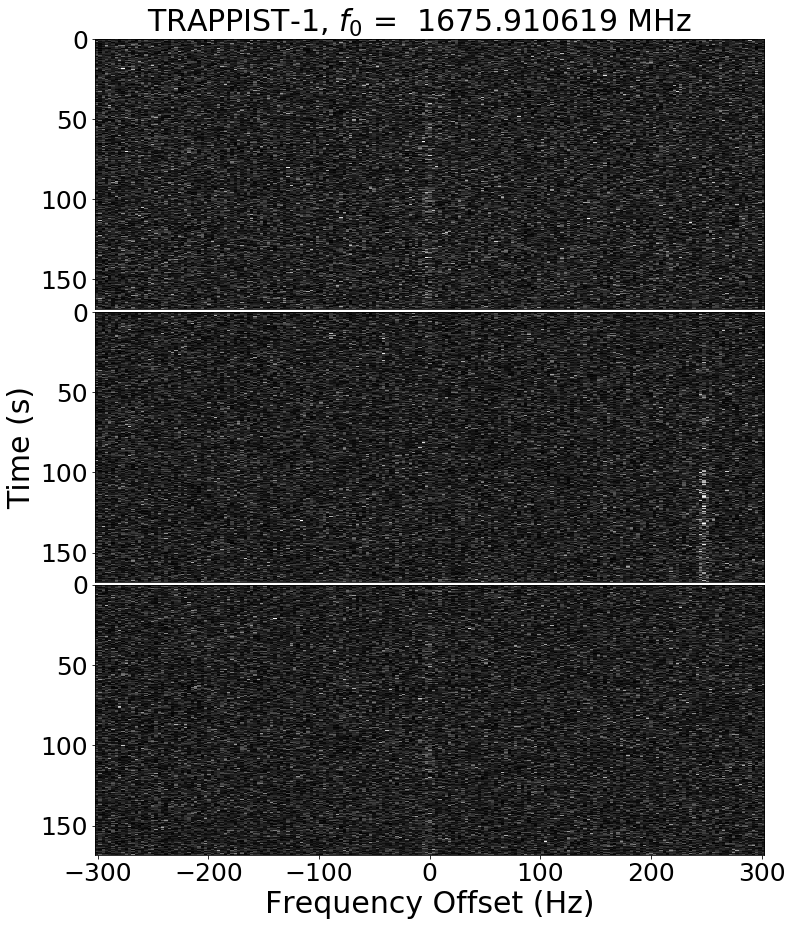}{0.3\textwidth}{}
          }
\caption{Time-Frequency diagrams for the ``on'' - ``off'' - ``on'' scans for sample category `f' candidates. The signal found during the Kepler-296 scan (left) has a time-frequency structure reminiscent of the ARSR structure.
  \label{fig:cat_f}}
\end{figure*}

Of the remaining 17 signals, 
we believe that 15 are anthropogenic.  These signals appear in three sources: Kepler-296 (3), LHS 1140 (5), and TRAPPIST-1 (7).  In each source, the signals exhibit similar modulation properties and are detected at similar frequencies but at different drift rates, implying that they are unlikely to be of extraterrestrial origin.
These signals are labeled as Category `f' in Table \ref{tab:final_cands}. The time-frequency diagrams of the ``on'' - ``off'' - ``on'' scans for a sample signal from each of the three sources are shown in Figure \ref{fig:cat_f}.

\begin{figure}[ht!]
\plottwo{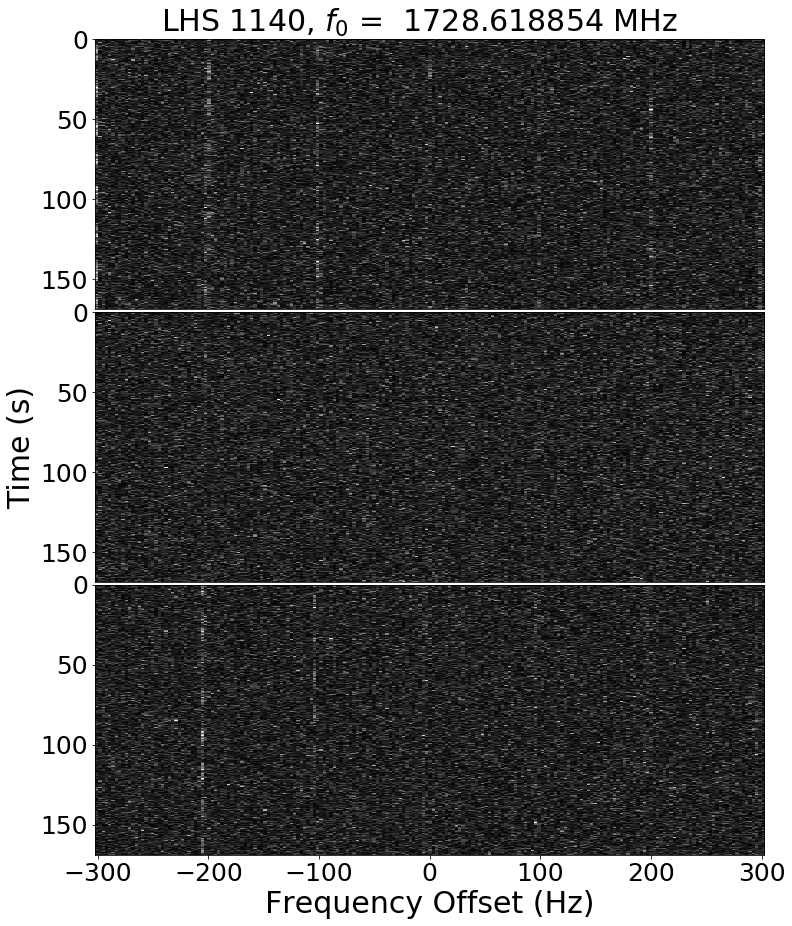}{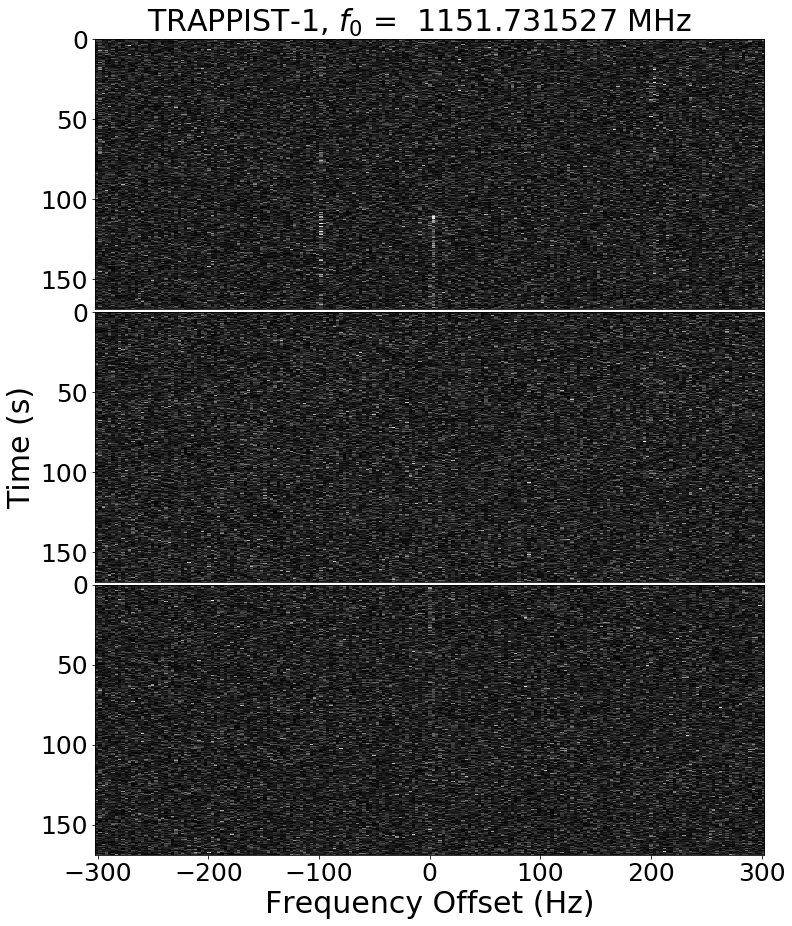}
\caption{Time-Frequency diagrams for the ``on'' - ``off'' - ``on'' scans for the 
most promising
technosignature candidates.  The candidate signal is at the center of each diagram. 
  Both signals have near-zero Doppler drift rate, and are in the vicinity of other signals that were discarded as anthropogenic RFI. For this reason, we cannot conclude that any
  extraterrestrial technosignatures were found in this search.  \label{fig:final_2}}       
\end{figure}

The remaining two candidates are shown in Figure \ref{fig:final_2}. Note that the drift rate for both candidates is close to zero
\added{(Table \ref{tab:final_cands})}
. Additionally, both candidates are found 
\replaced{neighboring}{within 100 Hz of}
other candidates flagged by our pipeline as RFI. As a result, we cannot conclude that these signals, nor any of the other signals found in this search, are of extraterrestrial origin.    

\section{Discussion}\label{sec:discussion}

\subsection{Drake Figure of Merit} \label{subsec:DFM}

The Drake Figure of Merit \citep{DFM} is often used to compare the parameter space examined by different searches and is given by 
\begin{equation} \label{eq:DFM}
DFM = \frac{\Delta \nu_{tot} \Omega}{S_{min}^{3/2}} 
\end{equation}
where $\Delta \nu_{tot}$ is the total bandwidth observed, $\Omega$ is the total angular sky coverage, and $S_{min}^{3/2}$ is the minimum flux density required for a detection. For our search, $S_{min}^{3/2} = 9.4$ Jy and $\Omega = N \times 0.015$ deg$^2$, where $N = 12$ is the number of
individual sky pointings \citep{Margot2018}. To compute $\Delta \nu_{tot}$, we take the
bandwidth of the GBT
L-band receiver (580 MHz)
and subtract the bandwidth of the GBT notch filter (141.2 MHz),
the bandwidth discarded due to known interferers (Table \ref{tab:known_RFI}; 137.167 MHz),
and the total bandwidth discarded during our density thresholding procedure (Section \ref{subsec:Freq_filters}; 37.101 MHz). Using the resulting bandwidth 
$\Delta \nu_{tot} = 264.532$ MHz,
we find that the DFM associated with this search is 
$1.6 \times 10^6$. 
This number amounts to about 1.7\% and 
10\% of the recent large surveys presented by \citet{Enriquez2017} and \citet{Harp}, respectively.

\subsection{Increase of Candidate Detection Efficiency} \label{subsec:cand_detection_improv}

The candidate detection procedures presented by \citet{Siemion2013, Enriquez2017, Margot2018} leave substantial regions of the spectrum unexamined, as pointed out in Section \ref{subsec:blanking}. This deficiency leads to an overestimation of the Drake Figure of Merit (DFM) associated with these searches, since this number is directly proportional to the total bandwidth examined (Equation \ref{eq:DFM}). 

We can calculate the
magnitude of this overestimation for the work of
\citep{Margot2018} by noting that a signal
with a drift rate $\dot{f}_i$ occupies no more than $\dot{f}_i \tau$ Hz of bandwidth, where $\tau$ is the scan duration. However, a window $f_w \approx 3000$ Hz was discarded around every detection, leaving $f_w - \dot f_i \tau$ Hz unexamined around every
candidate signal. We can calculate the approximate fraction $F$ by which the DFM was overestimated using 
\begin{equation}
F = \frac{N \Delta f_{tot}}{N \Delta f_{tot} -  \sum_i (f_w - \dot f_i \tau)}, 
\end{equation}
where $\Delta f_{tot}$ is the total bandwidth that was searched (300 MHz), and N is the total number of scans.
Using the database of detections found by \citet{Margot2018}, we find 
$F = 1.048$,
meaning
that their reported DFM was overestimated by 
approximately 5\%.
\citet{Enriquez2017} eliminate $f_w = 1200$ Hz of frequency space in their ``on'' scans centered around detections in their ``off'' scans.  In the absence of complete information about their detections, we are unable to compute the magnitude by which their DFM was overestimated.  

Our current procedure greatly improves on this situation by only requiring that detections do not overlap in time-frequency space. This improvement
allows us to search
the space around detected signals instead of discarding
frequency windows that are 1200-3000 Hz wide.  
As a result, our calculation of the DFM (Section \ref{subsec:DFM}) no longer suffers from
overestimation problems.  

\begin{figure}[ht!]
\plotone{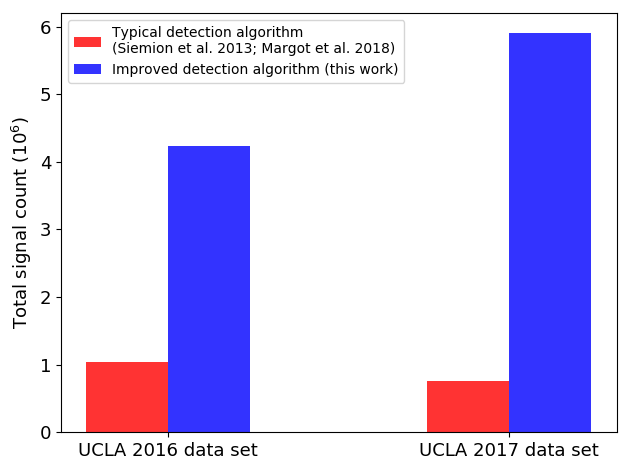}
\caption{
  Detection counts obtained with the algorithm of \citet{Siemion2013, Margot2018} (red) and that presented in this work (blue).
  The removal of many legitimate detections by typical algorithms suggests that claims of existence limits based on the results of these algorithms and others like it \citep[e.g.,][]{Enriquez2017} are questionable. \label{fig:signal_counts_compare}}       
\end{figure}

We can get a better idea of the level of improvement to our data processing pipeline by
comparing the signal counts recovered by
the candidate detection
method used by \citet{Margot2018} and 
the method that is used in this work (Figure \ref{fig:signal_counts_compare}).
Evidently, our improved detection procedure recovers more than four times the number of
detections reported by \citet{Margot2018}.
\added{Furthermore, 91\,283 of the 97\,083 signals that passed our Doppler and Direction-of-Origin filters (Section \ref{subsec:SQL_filters}) were only detected as a result of this improvement. Eight of our final 30 candidates (Table \ref{tab:final_cands}) would not have been detected with the detection algorithms of \citet{Margot2018}.}

The practice of blanking frequency space around candidate detections suggests  that a large number of candidates, which could include valid technosignatures, are removed from consideration when using the detection algorithms of \citet{Siemion2013, Enriquez2017, Margot2018}. This practice makes the calculation of existence upper limits unreliable, because these pipelines remove from consideration many of the signals that they are designed to detect.

We have reprocessed the 2016 data discussed by \citet{Margot2018} with our improved algorithms (Appendix \ref{sec:2016Results}). We found no evidence of extraterrestrial technosignatures.

\subsection{Existence Limits}

Considering that radio SETI detection pipelines typically eliminate substantial fractions of the spectrum (e.g., \citealp{Harp}, Table 2; \citealp{Enriquez2017}, 1200-1350 MHz; \citealp{Margot2018}, Table 2; this work, Table 2) and further eliminate a fraction of the signals that they are designed to detect \citep{Siemion2013, Enriquez2017, Margot2018}, it is difficult to make general and robust statements about the prevalence of narrowband emitters in the Galaxy. One such claim by \citet{Enriquez2017} has been shown to be questionable \citep{Margot2018}.
Injection of artificial signals in the data would demonstrate that a fraction of detectable and legitimate signals are not identified by existing pipelines.  Until this fraction is properly quantified, it is wise to refrain from making overly confident claims about the prevalence of radio emitters in the galaxy.   

\subsection{Sensitivity}
\citet{Margot2018} provide a detailed analysis of the sensitivity of a search performed with the 100 m GBT at a frequency resolution of $\Delta \nu = 2.98$ Hz sensitive to flux densities of 10 Jy.
The results of that calculation \citep[][Figure 5]{Margot2018} are
generally applicable here
because our search parameters are identical except for a slightly better sensitivity of 9.4 Jy.
We estimate that civilizations located near the closest of our observed sources (TRAPPIST-1; $\sim40$ ly) would require a transmitter with only a small fraction ($<1\%$) of the effective isotropic radiated power (EIRP) of the Arecibo planetary radar to be detectable in our search. Transmitters located as far as our most-distant observed source (Kepler-452; $\sim1800$ ly) require approximately 18 times the Arecibo EIRP. 

\section{Conclusions}\label{sec:conclusions}

We described the results of a search for technosignatures using two hours of GBT telescope time in 2017. We identified \totcount candidate signals, 98\% of which were automatically eliminated by our rejection filters. Of the signals that remained, \finalcountnoRFI were found outside of densely populated frequency regions and required further inspection. None of the remaining candidates were attributable to extraterrestrial technosignatures.

We found that quiet parts of the radio spectrum remain unexamined in the radio technosignature search pipelines of \citet{Siemion2013, Enriquez2017, Margot2018}.  This problem results in inflated estimates of figures of merit and unreliable upper limits on the prevalence of technosignatures.  To address this problem, we implemented a new procedure that increased the candidate detection efficiency by a factor of four or more compared to \citet{Margot2018}.

Our observations were designed, obtained, and analyzed by students enrolled in a UCLA course offered annually since 2016.  The search for technosignatures provides a superb educational opportunity for students in astrophysics, computer science, engineering, mathematics, planetary science, and statistics. In this work, two graduate students and 15 undergraduate students at UCLA learned valuable skills related to radio astronomy, telecommunications, programming, signal processing, and statistical analysis. A course syllabus and narrative is available at 
\url{http://seti.ucla.edu}.

\acknowledgments
We thank Michael Thacher and Rhonda Rundle, Arnie Boyarsky, Larry Lesyna, and David Saltzberg for the financial support that made the 2017 observations and analysis possible.
We thank the 2016 UCLA SETI class for providing a capable data-processing pipeline to build on.  We thank Smadar Gilboa, Marek Grzeskowiak, and Max Kopelevich for providing an excellent computing environment in the Orville L. Chapman Science Learning Center at UCLA. We are grateful to Wolfgang Baudler, Paul Demorest, John Ford, Frank Ghigo, Ron Maddalena, Toney Minter, and Karen O’Neil for enabling the GBT observations. \added{We are grateful to the reviewer for useful comments.}  The Green Bank Observatory is a facility of the National Science Foundation operated under cooperative agreement by Associated Universities, Inc. This research has made use of the NASA Exoplanet Archive, which is operated by the California Institute of Technology, under contract with the National Aeronautics and Space Administration under the Exoplanet Exploration Program. 

\vspace{5mm}
\facilities{Green Bank Telescope}

\newpage
\appendix

\section{Candidate Signal Detection and Bandwidth Estimation}\label{sec:bw}

To identify promising candidate signals in the drift-rate-by-frequency array output by the Doppler de-smearing algorithm, we applied an iterative procedure that seeks out candidate signals with SNR's exceeding $10$.
We began by arranging the $1023 \times 2^{20}$ array so that the drift rates decrease linearly down the rows of the array (i.e., the first row contains the integrated power
with drift rate 8.8644 \Hzsns, the second row contains the integrated power
with drift rate 8.8471 \Hzsns, and so on). We then searched for the highest SNR signal in this array, and noted its drift rate $\dot f_0$ and frequency $f_0$ at the start of the scan. 

\begin{figure}[ht!]
  \plotone{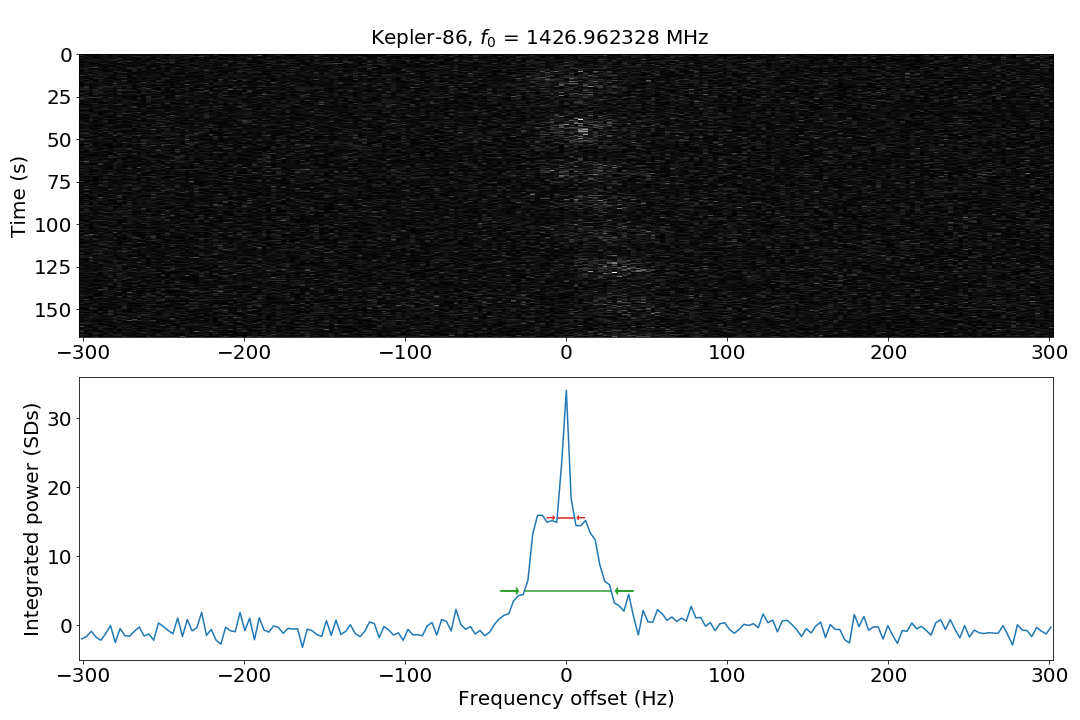}
  \caption{Sample result of the bandwidth estimation procedure. (Top) Time-Frequency diagram of signal. (Bottom) The power of the signal integrated
    with the best-fit drift rate of 0.2429 \Hzsns,
    in units of standard deviations of the noise ($\sigma$).
    The estimated bandwidth measured at 
FWHM
is shown by the red arrows
  and proved to be ineffective for the purpose of avoiding duplicate detections (e.g., signal outside this bandwidth exceeds our detection threshold of 10$\sigma$). 
  The green arrows show the bandwidth measured at 5$\sigma$,
  which yields a more robust estimate of the bandwidth and which we used in our implementation.
  \label{fig:bw}}
\end{figure}

In order to remove redundant detections, we used Equation \ref{eq:blanking}, which requires an estimate of the bandwidth of the detected signal.
We measured the bandwidth of candidate signals with the Python SciPy routine {\tt peak\_widths}.
These routines measure the width from the central peak of the signal to the first point on either side above a pre-defined threshold value, using linear interpolation when necessary.
We initially measured the bandwidth  as the 
full width at half maximum (FWHM)
of each signal.
However, due to the  peculiar nature of many of the detected signals
this threshold value proved to be ineffective.
We instead set the bandwidth measurement threshold at
five times the standard deviation of the noise, 
which corresponds to half of our signal detection threshold (Figure \ref{fig:bw}).

By default, we used a search window of 200 frequency bins
($\sim600$ Hz)
on either side of
each signal
to measure the bandwidth.
For narrowband signals, this range is more than enough to ensure that the entirety of a signal fits within the search window. If a detected signal is close to the edge of the channel and the required frequency window is not available on one side of the signal, we utilize the entire available frequency range on that side of the signal, and the full 600 Hz on the other. 
The SciPy routines conveniently provide the frequency coordinates of the left and right intersection points at the specified threshold, which we used to calculate the width of the signal to the left ($\Delta f_{b_l}$) and right ($\Delta f_{b_r}$) of the center frequency.

Signals with large bandwidths required additional care.  If the median integrated power of the signal within the initial 1200 Hz window exceeded $5\sigma$, we labeled the signal as broadband and increased the search window to 200000 frequency bins, or $\sim 600$ kHz, on either side of the main signal. We applied a Savitzky-Golay filter \citep{SGfilter} with a window of 1001 frequency bins ($\sim 3000$ Hz) and a polynomial order of 3 to the integrated spectra within the search window. This filter reduces noise by fitting a polynomial to all the points within the specified window and replacing the central point with the corresponding fit value.  This filter was chosen for its computational advantages and simplicity.
We then applied the bandwidth estimation procedure described above to the filtered points and stored the result in our database.

We used the bandwidth measurements in Equation \ref{eq:blanking}, as follows:
\begin{eqnarray} \label{eq:blankingA}
f_0 - \Delta f_{b_l}  <  f <  f_0 + (\dot f - \dot f_0 ) \tau + \Delta f_{b_r} \qquad \text{if} \quad \dot f > \dot f_0 \\
\nonumber
f_0 + \Delta f_{b_r}  >  f >  f_0 + (\dot f - \dot f_0 ) \tau - \Delta f_{b_l}  \qquad \text{if} \quad \dot f < \dot f_0
\end{eqnarray}
However, we chose to establish a minimum bandwidth to account for
uncertainty in the determination of the drift rate of a signal (for example, if a signal's drift rate is time variable or is not a perfect multiple of 
the drift rate step $\Delta \dot{f}$).
Specifically, we 
redefined $\Delta f_{b_l}$ as 
min($\Delta f_{b_l}$, $3\Delta \nu$)
and
$\Delta f_{b_r}$ as 
min($\Delta f_{b_r}$, $3\Delta \nu$), which amounts to
a minimum of
$\sim 10$~Hz on either side of each signal, for a minimum of 20 Hz per signal. 

\begin{figure}[htb]
\plotone{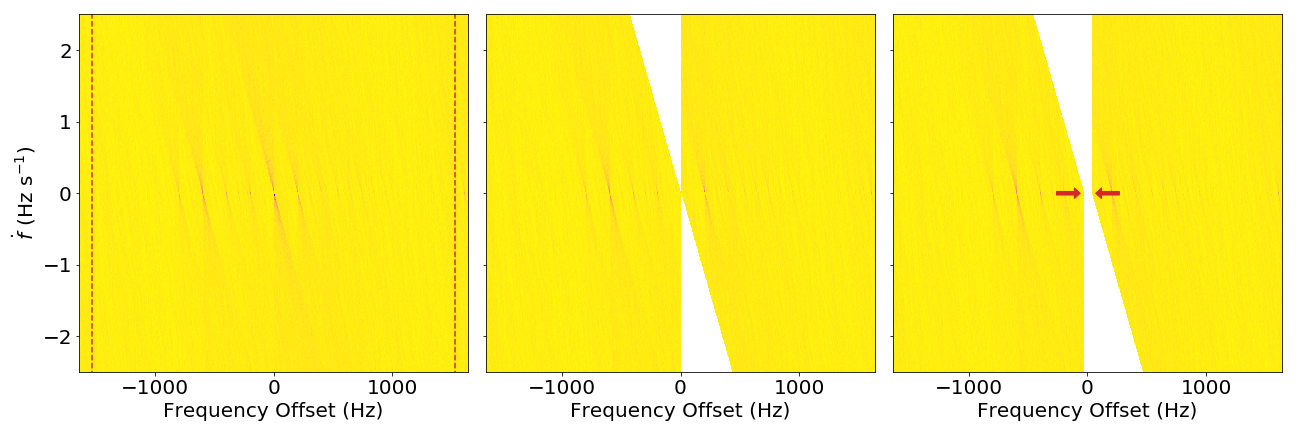}
\caption{(Left) A portion of the
  original array output by the Doppler de-smearing algorithm for a sample scan (LHS 1140). The plot is centered on $f = 1546.879303$ MHz. The intensity of the plot represents the integrated power at a given drift rate and frequency.
  Signals of interest are represented as local maxima. The minimum integrated power in this array exceeds our 10 SNR detection threshold, therefore all signals shown are detectable by our pipeline.
   The
signal with  
  maximum SNR 
  in this portion of the array 
  is located at the center 
frequency.  
  The area between the dashed vertical lines represents the slice of the array that would have been discarded after a signal detection with the approach of \citet{Siemion2013} and \citet{Margot2018}, i.e., only one of the 
  signals would have been reported. (Middle) Same array after application of our new masking procedure, assuming that the measured bandwidth of the signal is 0.
  With our new procedure, 
  valid signals in the vicinity of the strongest signal are not discarded.
  (Right) Same array after application of our new masking procedure, assuming that the measured bandwidth of the signal is 60 Hz.
  This number was chosen for visualization purposes, and does not represent the true measured bandwidth
($\sim3$ Hz)  
   of the center peak.
  \label{fig:masking}
}
\end{figure}

We removed redundant detections with frequencies given by Equation \ref{eq:blankingA} by applying a mask to the drift-rate-by-frequency array. This mask is defined by drawing two lines through the array. The first is a vertical line at the frequency of the detected candidate signal. The second is a line with a slope 
of $- \Delta \dot f / \Delta \nu$
crossing through the frequency and drift rate point corresponding to the detected signal (Figure  \ref{fig:masking}). The boundaries defined by these two lines comprise the mask for zero bandwidth signals. In order to account for the finite bandwidth of a signal, the boundaries are shifted to the left and right so that the total width of the mask at the detection point matches the measured bandwidth of the signal (signified by the red arrows in Figure \ref{fig:masking}). 

\section{Doppler and Direction-of-Origin Filters} \label{sec:SQLfilters}

To distinguish anthropogenic signals from potential extraterrestrial technosignatures, we
invoked several filter procedures within our database to flag promising technosignature candidates. The first filter flags candidate signals with non-zero Doppler drift rates.

The direction-of-origin filters require signals from different scans to be compared and possibly paired. We pair two signals if they have similar drift rates and
compatible frequencies, i.e., the frequencies at the beginning of each scan are within some tolerance of a linear relationship with a slope equal or nearly equal to the drift rate.
We quantify these tests as follows. Consider a signal with start time $t_0$, start frequency $f_0$, and drift rate $\dot f_0$, and another signal from a different scan with corresponding values $t$, $f$, $\dot f$.  We define $\Delta t = t - t_0$ and require that $f$ falls in the interval $[f_-,f_+]$ for pairing, where
\begin{equation}
f_{\pm} = (\dot f_0 \pm 2\Delta \dot{f}) \Delta t  \pm \Delta \nu.
\end{equation}

In this work, the values $\Delta \dot{f}$ and $\Delta \nu$
are given by 0.0173 \Hzsns and 2.98 Hz, respectively.
To account for uncertainty in the drift rate determination, we allow for a drift rate difference of 
$\Delta \dot{f}$.
We thus query the database for all candidate signals
with a frequency at the start of the scan 
in the range $[f_-, f_+]$ with a drift rate 
$\dot f_0 - \Delta \dot{f} \leq \dot f \leq \dot f_0 + \Delta \dot{f}$ \Hzsns. 
Two signals are considered paired if the following condition holds:
\begin{equation}\label{eq:SQLcrit}
\min_i|(f - f_0) + \dot f_i \Delta t| \leq \Delta \nu
\end{equation}
where $\dot f_i \in \{\dot f_0, \dot f_0 \pm \Delta \dot{f}, \dot f, \dot f \pm \Delta \dot{f}\}$.

To determine whether a signal is persistent, i.e., whether it is detected in both scans of its source, 
we apply the above procedure to each candidate and all candidates detected in the second scan of the same source. If a match is found, both signals are flagged. To 
determine whether a signal's direction of origin is unique, i.e, to ensure that it is not detected in the scans corresponding to other sources, we apply the above procedure to each candidate and all candidates detected in either scan of all other sources.  All matches are discarded from consideration.  The candidate signals that remain are flagged as having a single direction of origin.

\section{Signal Density Thresholding} \label{sec:DensityThresh}

\begin{figure}
  \plotone{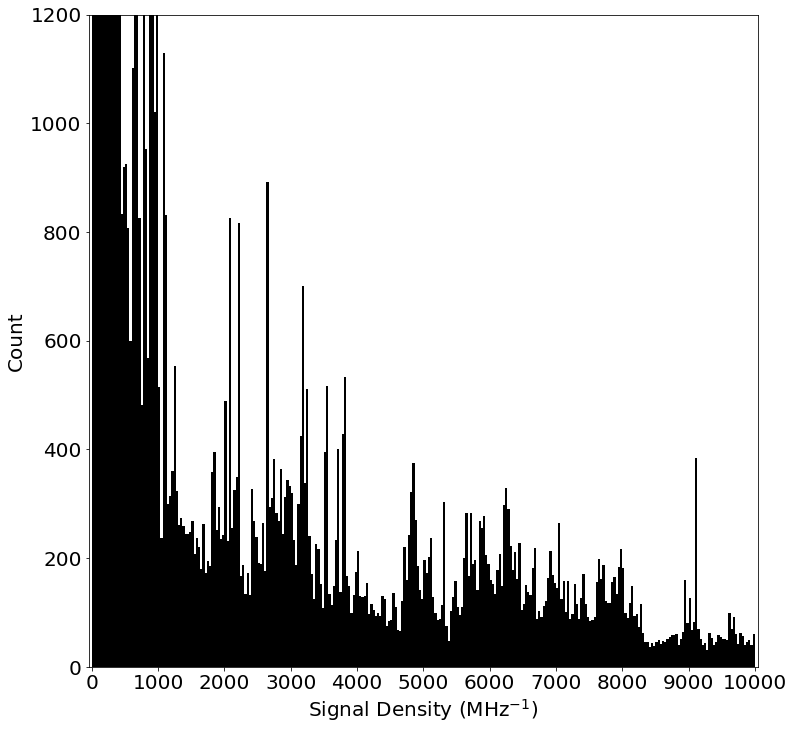}
\caption{Signal densities of 1kHz regions in the 1.15--1.73 GHz range. The plot is clipped at a max of 1200 signals MHz$^{-1}$.
  Note the sharp dropoff at approximately 1000 signals MHz$^{-1}$. \label{fig:density_hist}}
\end{figure}

In order to remove candidate signals that were likely to be
anthropogenic RFI, we developed a procedure to filter signals based on the
density of nearby detections. We began by dividing the
1.15 -- 1.73 GHz range into 1-kHz-wide frequency bins. For each bin, we measured the signal density by counting the number of
detections within a window centered on the bin.
For these calculations, we excluded
the regions listen in Table \ref{tab:known_RFI}. We tested four different window sizes: 1 kHz, 10 kHz, 100 kHz, and 1 MHz. For each window size, we plotted a histogram of
signal counts.
We found that a 1 MHz window resulted in a distinctive transition between small and large signal densities at approximately 1000 signals MHz$^{-1}$ (Figure \ref{fig:density_hist}), and we used this threshold and window size to filter out regions of high signal density.

\section{Re-analysis of 2016 Data} \label{sec:2016Results}

\citet{Margot2018} presented the results of a search for technosignatures around 14 planetary systems in the Kepler field conducted on April 15, 2016, 16:00 - 18:00 universal time (UT) with the GBT.  They discarded a frequency region of width $\pm \dot{f}_{max} \tau \simeq 3000$ Hz around each detection, where $\dot{f}_{max} = 8.86$ \Hzs is the maximum drift rate detectable by their search and $\tau \approx 150s$ is the integration time of the scan.  This process or a variant of it is
also implemented in the radio SETI detection pipelines of \citet{Siemion2013} and \citet{Enriquez2017}.  As a result, a substantial portion of the spectrum remains unexamined in these searches, and the Drake figure of merit associated with these searches is overestimated (Section \ref{sec:discussion}). 

To
remedy
this situation, we have re-analyzed the data obtained
by \citet{Margot2018}
using the candidate detection procedure described in this work. We have found a total of 4\,228\,085 signals as compared with 1\,046\,144 reported previously. Our
Doppler and direction-of-origin rejection algorithms (Section \ref{subsec:SQL_filters}, Appendix \ref{sec:SQLfilters}) automatically labeled more than 99\% of the detected signals as anthropogenic RFI. After removing all remaining signals found within operating bands of the interferers described by
\citet[][Tables 2 and 3]{Margot2018},
we were left with 18
technosignature candidates. Seven of these
had been identified by \citet{Margot2018} and attributed to anthropogenic RFI.  
The properties of the remaining 11 technosignature candidates are given in Table \ref{tab:final_cands_2016}. These signals were further scrutinized and categorized according to the procedure described in Section \ref{sec:results}. All were found to be attributable to anthropogenic RFI. 

We note that 
3
of the 19 final technosignature candidates described
by  \citet{Margot2018} 
were 
found to be part of a broadband RFI signal and were removed via Equation \ref{eq:blanking}.
The remaining 9 signals were correctly
labeled as RFI by our Doppler and direction-of-origin rejection algorithms or frequency-based filters. This enhancement in classification performance was only possible because of the improvements to the candidate signal detection algorithms presented in this work
and because the raw data were preserved for re-analysis.

\begin{deluxetable}{lcCccc}
\caption{Characteristics of Top 11 Candidates from 2016 Search
\label{tab:final_cands_2016}}
\tablehead{
\colhead{Source} & \colhead{Epoch (MJD)}  & \colhead{Frequency (Hz)} & \colhead{Drift Rate \Hzs} & \colhead{SNR} & \colhead{Category}}
\startdata
Kepler-22  & 57493.70935185 & 1456613901.257515 &  0.52042 &  14.3 & a \\ 
\hline
Kepler-296 & 57493.71608796 & 1454808911.681175 &  0.05204 &  10.3 & a \\ 
           &                & 1454877239.465714 &  0.01735 &  14.0 & a \\ 
           &                & 1457440766.692162 &  0.31225 &  12.9 & a \\ 
           &                & 1457443791.627884 &  0.31225 &  11.9 & b \\ 
\hline
Kepler-399 & 57493.68870370 & 1414058676.362038 &  0.01735 & 764.4 & b \\ 
\hline
Kepler-407 & 57493.71819444 & 1620381447.672844 & -6.40113 &  14.5 & d \\ 
\hline
Kepler-440 & 57493.74275463 & 1457486632.466316 &  0.41633 &  21.4 & b \\ 
           &                & 1457486835.122108 &  0.39899 &  23.6 & b \\ 
\hline
Kepler-442 & 57493.74497685 & 1308946093.916893 & -0.01735 &  11.3 & e \\ 
           &                & 1376548311.114311 & -0.24286 & 540.9 & b \\ 
\enddata
\tablecomments{
  Properties are listed for the first scan of the source only. For a description of the `Category' column, see Section \ref{sec:results}.}
\end{deluxetable}

\listofchanges

\end{document}